\documentclass{ifacconf}
\usepackage{amsmath}
\usepackage{amsfonts}
\usepackage{amssymb}
\usepackage{mathtools}
\usepackage{color}
\usepackage{graphicx}      
\usepackage{natbib}        
\newcommand\RE{\mathbb{R}}

\newcommand\mymatrix[2]{\left[\begin{array}{#1} #2 \end{array}\right]}

\DeclareMathOperator{\atanh}{arctanh}

\newtheorem{defi}{Definition}
\newenvironment{definition}{\begin{defi}\rm }{\hfill \hspace*{1pt} \hfill $\lrcorner$\end{defi}}
\begin{document}
\begin{frontmatter}
\title{Differentially passive circuits \\ that switch and oscillate
} 
\thanks[footnoteinfo]{The research leading to these results has received funding from the European Research Council under the Advanced ERC Grant Agreement Switchlet n. 670645.}
\author[All]{F.A. Miranda-Villatoro} 
\author[All]{F. Forni} 
\author[All]{R. Sepulchre}
\address[All]{Department of engineering, University of Cambridge, UK
\\ {\small(e-mail: fam48@cam.ac.uk, f.forni@eng.cam.ac.uk, r.sepulchre@eng.cam.ac.uk)}.}
\begin{abstract}                
The concept of passivity is central to analyze circuits as interconnections of  passive components.  We illustrate that when used {\it differentially}, the same concept leads to an interconnection theory for  electrical circuits that switch and oscillate  as interconnections of passive components with operational amplifiers (op-amps). The approach builds on recent results on dominance and $p$-passivity aimed at generalizing dissipativity theory to the analysis of 
non-equilibrium nonlinear systems. Our paper  shows how those results apply to basic and well-known nonlinear circuit architectures. They illustrate the potential of dissipativity theory to design and analyze switching and oscillating circuits quantitatively, very much like their linear counterparts.
\end{abstract}
\end{frontmatter}
\section{Introduction}
The concept of passivity originates in circuit theory. It characterizes circuit elements that can possibly store
and dissipate the energy provided by the environment, but not the other way around. 
Passivity is inherently an interconnection concept:  passive interconnections
of passive components model passive circuits. Dissipativity theory, the system theoretic generalization of passivity theory,
has become a cornerstone of system theory. It provides an interconnection theory to design and analyze
stable dynamical systems. Such systems dissipate the energy stored internally and  provided externally.
In short, dissipativity theory is  an interconnection theory for Lyapunov stability analysis.
In recent years, many researchers have pointed to the relevance of studying stability {\it incrementally}
or {\it differentially} when addressing questions that go beyond the stability analysis of isolated equilibria. 
Differential stability concepts include contraction theory \citep{lohmiller1998},
convergence theory \citep{pavlov2005}, or differential
Lyapunov theory \citep{forni2014b}. They have proven relevant in a number of areas, most prominently in questions
pertaining to nonlinear observers \citep{aghannan2003}, oscillator synchronization \citep{stan2007},
or regulation theory \citep{jouffroy2010}.
Differential dissipativity is to differential stability what dissipativity is to Lyapunov stability. 
It was introduced in the recent papers \citep{forni2013, forni2013b, schaft2013}.
The present paper aims at illustrating the potential of  differential passivity as an interconnection theory 
of circuits that can switch an oscillate. In contrast with classical dissipativity theory, such a theory must cope 
with the stability analysis  of dynamical systems
that posses multiple  equilibria or limit cycles. It is based on the concept of $p$-dominance and $p$-passivity 
recently introduced in \cite{Forni2017b, forni2017}. Intuitively, the attractors of a $p$-dominant system are the
attractors of a $p$-dimensional system: a unique equilibrium for $p=0$, but possibly multiple equilibria for $p=1$,
and limit cycles for $p=2$. We interpret classical differential dissipativity theory as an interconnection theory for $0$-dominance, that is, differential stability. 
The extension to  $p=1$ and $p=2$ is motivated by the analysis of multistability or limit cycles in interconnected systems.
Nonlinear circuit theory provides a realm of switching and oscillatory behaviors designed
from the simple  elements of linear circuit theory interconnected with operational amplifiers (op-amp) \citep{clayton2003operational}.
Our aim in the present paper is to illustrate that those building blocks are the  natural building blocks of $p$-passive circuits.
System theoretic tools have lacked so far for the quantitative analysis and design of  such circuits. Their analysis
normally rests on simplifying assumptions, time-scale separation arguments leading to asymptotic
analysis, or reductions to two-dimensional phase portraits.   In contrast, we are aiming at 
quantitative and  computationally tractable certificates such as those
used in the theory of linear time-invariant systems. Such tools have made the success of
 robustness and performance analysis of linear time-invariant systems.
 A pillar of passivity theory is the passivity theorem, which states that the
 negative feedback interconnection of two passive systems is passive. It
 is also well-known that  switches and oscillatory circuits require both positive and negative
 feedback interconnections of op-amps with passive elements. We stress in the present paper that
 such interconnections fall in the category of the $p$-passivity theorem, which states that
 the negative feedback interconnection of a $p_1$-passive with a $p_2$-passive circuit
 is $p_1+p_2$ passive. 
 
The differential analysis in this paper assumes smooth systems. A companion paper shows
 that an analog framework exists for non-smooth systems.  To account for the lack of differentiability,
 {\it differential} concepts have then to be replaced by {\it incremental} concepts. Many of the nonlinear circuits discussed in the present paper
 have a non-smooth analog that falls in the category of linear complementarity systems studied in \cite{miranda2018}.
The rest of the paper is organized as follows.
Section \ref{sec:dom_and_pass} provides a brief summary of the concepts of dominance and  $p$-passivity.
Section \ref{sec:operational_amplifier} revisits standard properties  of the operational amplifier   its differential passivity properties. 
Section \ref{sec:switch_and_osc} revisits the  basic architectures of circuits that switch and oscillate, analyzing those systems
as both positive and negative feedback interconnections of operational amplifiers with  passive linear circuits. The discussion in
Section \ref{sec:mixedfeedback_and_modulation} suggests that those architectures are robust and amenable to regulation. 
 
\section{Dominance and differential passivity}
\label{sec:dom_and_pass}
We consider the nonlinear system 
\begin{equation}
\label{eq:csys}
\dot{x} = f(x) 
\end{equation}
where $x \in \RE^n$ and $f$ is a smooth vector field.
The \emph{prolonged system} consists of
\eqref{eq:csys} augmented with the linearized equation
$
\dot{\delta x} = \partial f(x) \delta x
$,
where $\partial f(x)$ denotes the Jacobian linearization of $f$.
By construction $\delta x \in \RE^n$ (identified with the tangent space of $\RE^n$).
An important notion for this paper is the  inertia $(p,0,n-p)$ of a symmetric
matrix, meaning that the matrix has $p$ eigenvalues in the open left half-plane, $0$ eigenvalues
on the imaginary axis, and  $n-p$ eigenvalues in the right half-plane.  
The following definition is taken from \citep{Forni2017b}.
\begin{definition}
\label{def:diff-p-dominance}
A nonlinear system \eqref{eq:csys} is \emph{$p$-dominant 
with rate $\lambda\geq 0$} if there exist a 
constant symmetric matrix $P$ with inertia $(p,0,n-p)$
and $\varepsilon \geq 0$ for which the prolonged system 
satisfies
\begin{equation}
\label{eq:p-dominance}
\mymatrix{c}{\!\dot {\delta x}\! \\ \!\delta x\!}^{\top} \!
\mymatrix{cc}{
0 & P \\ P & \ 2\lambda P + \varepsilon I
}
\mymatrix{c}{\!\dot {\delta x}\! \\ \!\delta x\!}
\leq 0
\end{equation}
for all $(x,\delta x)\in \RE^n \times \RE^n$ The property is strict if $\varepsilon>0$.
\end{definition}
Solving \eqref{eq:p-dominance} is equivalent to finding a uniform solution $P$ 
to the linear matrix inequalities 
$
\partial f(x)^{\top} P + P \partial f(x) + 2 \lambda P \leq -\epsilon I 
$
for all $x\in \RE^n$. For a linear system $\dot{x} = Ax$, the inequality  reduces
to $(A+\lambda I)^{\top} P + P (A+\lambda I) \leq -\epsilon I$,
which is feasible for linear systems whose eigenmodes can be split into
$p$ {\it dominant} modes and $n-p$ {\it transient} modes, 
separated by the rate $\lambda$, \citep[Proposition 1]{forni2017}. 
For nonlinear systems $p$-dominance captures 
the property that the asymptotic behavior of the system 
is  $p$-dimensional. This intuitive characterization is made precise in
the following
result  \citep[Corollary 1]{Forni2017b}.
 \begin{thm}
\label{thm:behavior}
Let  \eqref{eq:csys} be a 
strictly $p$-dominant system with rate $\lambda \geq 0$.
Then, every bounded solution of \eqref{eq:csys} asymptotically converge to
\begin{itemize}
\item a unique fixed point if $p = 0$;
\item a fixed point if $p = 1$;
\item a simple attractor if $p=2$, i.e. a fixed point, a set of fixed points and their connected arcs, or a limit cycle. 
\end{itemize}
\end{thm}
In what follows we will study $p$-dominant systems as
interconnections of open systems. We consider open systems of the form
\begin{equation}
\label{eq:osys}
\dot{x} = f(x) + B u \,, \quad y = Cx
\end{equation}
where $u \in \RE^m$ and $y \in \RE^q$ define 
the input and the output to the system, respectively. $B$ and $C$
are matrices of appropriate dimension.
The prolonged system to \eqref{eq:osys} is obtained by 
augmenting  \eqref{eq:osys} with the linearized equations
$ \dot{\delta x} = \partial f(x) \delta x + B \delta u $,
$ \delta y = C \delta x $. 
The following definition is taken from \citep{Forni2017b}.
\begin{definition}
\label{def:p-passivity}
A nonlinear system \eqref{eq:osys} is $p$-passive from $u$ to $y$
with rate $\lambda \geq 0$ if there exist a 
constant symmetric matrix $P$ with inertia $(p,0,n-p)$
and $\varepsilon \geq 0$ for which the prolonged system 
satisfies
\begin{equation}
\label{eq:diff-p-dominance}
\mymatrix{c}{\!\dot {\delta x}\! \\ \!\delta x\!}^{\top} \!
\mymatrix{cc}{
0 & P \\ P & \ 2\lambda P + \varepsilon I
}
\mymatrix{c}{\!\dot {\delta x}\! \\ \!\delta x\!}
\leq 
\mymatrix{c}{\!\delta y\! \\ \! \delta u\!}^{\top}
\mymatrix{cc}{
0 & I \\ I & 0
}
\mymatrix{c}{\!\delta y\! \\ \! \delta u\!}
\end{equation}
for all $(x,\delta x)\in \RE^n \times \RE^n$ and all $(u,\delta u)\in \RE^m \times \RE^m$.
The property is strict if $\varepsilon>0$.
\end{definition}
The concept of $p$-passivity is  related to $p$-dominance   in the same way as passivity is related to stability.
Differential \citep{forni2013b} or incremental \citep{pavlov2008} passivity are
synonyms of  $0$-passivity.
For a static differentiable nonlinearity $y = \varphi(u)$,
$0$-passivity simply means monotonicity, that is positivity of its derivative: 
if  $\partial \varphi(u) \geq 0$, then
$\delta y^T \delta u = (\partial \varphi(u)\delta u )^\top \delta u \geq 0$ for all $\delta u$. 
 
 The following $p$-passivity theorem is the natural extension of the classical passivity theorem.
It is taken  from \citep[Theorem 4]{Forni2017b}.
\begin{thm}
	\label{thm:interconnection}
	Let $\Sigma_{1}$ and $\Sigma_{2}$ be (strictly) $p_{1}$ and 
	$p_{2}$ passive, respectively, from input $u_i$ to output $y_i$, $i \in \{1,2\}$,
	both with rate $\lambda \geq 0$, 
	Then, the negative feedback interconnection 
	\begin{displaymath}
		u_{1} = -y_{2} + v_{1}, \quad u_{2} = y_{1} + v_{2}
	\end{displaymath}
	of $\Sigma_{1}$ and	$\Sigma_{2}$ is (strictly) $(p_{1} + p_{2})$-passive from
	$v = (v_{1}, v_{2})$ to $y = (y_{1}, y_{2})$, with rate $\lambda$. 
	The interconnection is (strictly) $(p_{1} + p_{2})$- dominant.
	\label{eq:dissipativeConexion}
\end{thm}
We observe that negative feedback preserves $p$-passivity only if the 
two components share a common rate $\lambda$. 
For linear systems of the form $\dot{x} = Ax + Bu$, $y=Cx$, 
$p$-passivity has a useful frequency domain characterization in terms of the 
shifted transfer function  $G(s - \lambda) = C (s I - (A + \lambda I))^{-1} B$,
as shown by the next theorem from \cite{Miranda2017b}.
\begin{thm}
	\label{prop:frequency}
	A linear system is $p$-passive with rate $\lambda$ if and only if the following two conditions hold,
	\begin{enumerate}
		\item\label{cond:1} $\Re \left\{ G(j \omega - \lambda) \right\} \geq 0$, 
			for all, $\omega \in \RE \cup \{+\infty\}$.
		\item\label{cond:2} $G(s - \lambda)$ has $p$ poles on $\mathbb{C}_{+}$.
	\end{enumerate}
	The property is strict if $G(s - \lambda)$ has $p$ poles in the interior of $\mathbb{C}_{+}$
\end{thm}
\section{The operational amplifier is $0$-passive}
\label{sec:operational_amplifier}
\begin{figure}[htpb]
	\centering
	\includegraphics[width = 1\columnwidth]{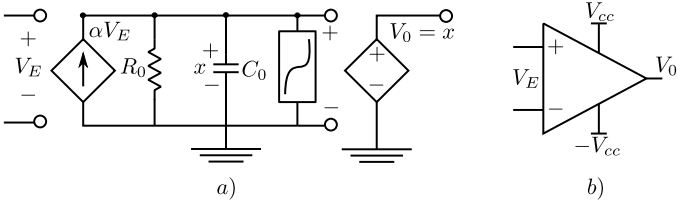}
	\caption{First order op-amp model with saturation: a) internal structure, b) symbolic representation.}
	\label{fig:opamp}
\end{figure}
\begin{figure}[htpb]
	\centering
	\includegraphics[width = 0.3\textwidth]{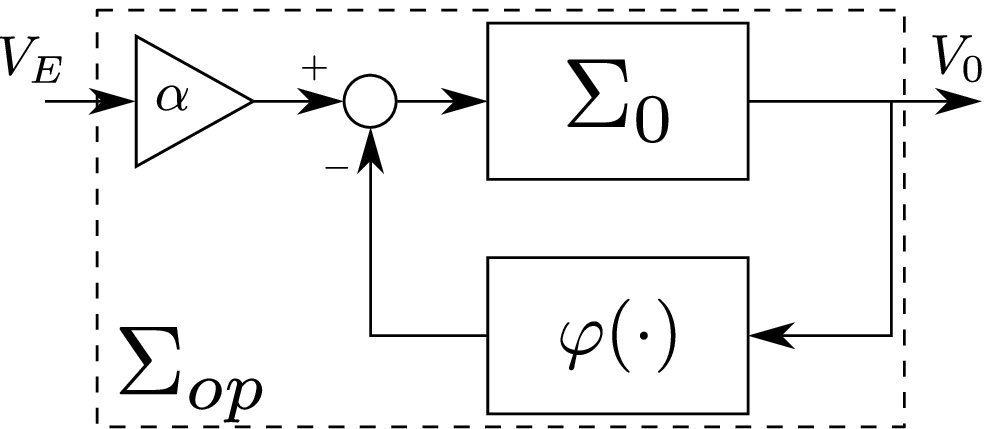}
	\caption{Open-loop operational amplifier model}
	\label{fig:lureSys}
\end{figure}

Figure \ref{fig:opamp} represents a classical model of operational amplifiers \citep{karki2000, noseek2009}:
the $RC$ network is a circuit realization of a first order
linear model in parallel with a voltage-controlled 
current source $\alpha V_{E}$, where $\alpha \in (0, +\infty)$, \citep{noseek2009}.
The $RC$ network models the finite bandwidth property of a real device and it is connected to a static
nonlinear element
\begin{equation}
i = \varphi(V)
\end{equation}
to account for the bounded range of $V_{0} = x$.
The static nonlinear element is a smooth and odd 
nonlinearity modelled for instance as follows:
\begin{equation}
	i = \varphi(V) := \eta  \sinh(\beta V)
	\label{eq:sn}
\end{equation}
where $\eta > 0$ and $\beta > 0$ are  suitable parameters.
The results
of this paper hold for any stiffening nonlinearity $\varphi$ 
satisfying the following assumption \citep{stan2007}. 
\begin{assum}\label{assum:nonlinear}
The static nonlinearity $\varphi: \RE \to \RE$ is an odd function 
such that $\frac{\partial \varphi(y)}{\partial y} \in [0, +\infty)$.
Furthermore, for any $k > 0$,  there
exists a $r > 0$ such that
\begin{equation}
	y \varphi(y) - k y^{2} > 0, \  \text{ for all } \vert y \vert > r \ .
	\label{eq:assum:nonlinear}
\end{equation}
\end{assum}
For example, $\varphi(y) = y^{2n + 1}$, for $n \in \mathbb{N}$, $\varphi(y) = \atanh(y)$
and $\varphi(y) = \sinh(y)$ all satisfy Assumption \ref{assum:nonlinear}.
The first-order model of the op-amp in Figure \ref{fig:opamp} 
has the state-space model
\begin{equation}
	\label{eq:op-amp}
	\Sigma_{op}:
	\begin{cases}
		\dot{x} = - \frac{1}{R_{0} C_{0}} x - \frac{1}{C_{0}}  \varphi 
		\left( x \right) + \frac{\alpha}{C_{0}} V_{E}
		\\
		V_{0} = x
		\end{cases}	
\end{equation}
which, notably, admits the block diagram representation of the 
Lur'e system in Figure \ref{fig:lureSys}. 
The transfer function of the linear part $\Sigma_0$ 
\begin{equation}
	G_{0}(s) = \frac{\frac{1}{C_{0}}}{s + \frac{1}{R_{0} C_{0}}}
\end{equation}
is a strictly $0$-passive network (by Theorem \ref{prop:frequency})
with rate $\lambda \in [0, \frac{1}{R_{0} C_{0}})$.
The op-amp is thus given by the negative feedback interconnection of a 
strictly $0$-passive linear system with a static $0$-passive nonlinearity
(under Assumption \ref{assum:nonlinear}). Therefore, by
Theorem \ref{thm:interconnection}, the closed loop is  
a strictly $0$-passive system from $V_E$ to $V_0$ 
with rate $\lambda \in [0, \frac{1}{R_{0} C_{0}})$. The same representation would hold
if the $RC$ circuit was replaced by any $0$-passive network.
\section{Switches and oscillators via feedback amplifiers}
\label{sec:switch_and_osc}
\subsection{Feedback and boundedness}
\label{sec:feedback_and_boundedness}
The great versatility of the op-amp comes from its interconnection properties.
The device allows for a wide range of behaviors, enabled
by the interconnection of the op-amp with suitable linear networks.
The circuits in this paper will only include interconnections of op-amps
with  linear
stable networks
\begin{equation}
	\dot{z} = A z + B u, \quad
	y = C z \ .
	\label{eq:linearNetwork}
\end{equation}
where $z \in \RE^{n}$, $u \in \RE$, $y \in \RE$ are  
the state, input, and output of a generic linear network,
respectively. $A$, $B$ and $C$ are 
constant matrices of appropriate dimensions. 
Such interconnections always lead to bounded behaviors:
\begin{thm}
	Suppose that  $A$ is  a Hurwitz matrix. Under Assumption \ref{assum:nonlinear}, the trajectories of the 
	system defined by \eqref{eq:op-amp},  \eqref{eq:linearNetwork}, and the interconnection rule
	\begin{equation}
		V_E = \pm Cz+V_r \quad u = V_{0}
	\end{equation}
 	are all bounded, for any constant voltage $V_r\in \RE$. 
	\label{thm:boundedness}
\end{thm}
\begin{pf}
	Let $V:\RE \times \RE^{n} \to \RE$ be the positive definite function 
	\begin{equation}
		\label{eq:lyap}
		V(x,z) = \frac{1}{2} x^{2} + \frac{1}{2} z^{\top} P z,
	\end{equation}
	where $P = P^{\top} > 0$ satisfies $A^{\top} P + P A\leq - Q$, for some 
	$Q = Q^{\top} > 0$. Then, taking $\eta = \frac{1}{R_0C_0}$, $\beta = \frac{1}{C_0}$,
	$\rho = \alpha \Vert {C} \Vert + 2 \Vert P {B} \Vert$, $\varepsilon > 0$, and
	$\mu_{Q}$ given by the smallest eigenvalue of $Q$, we have 
	\begin{align*}
		\dot{V}(x, z) &= -\eta x^{2} - z^{\top} Q z - \beta x \varphi(x) \pm \alpha x {C} z 
		+ 2 z^{\top} P {B} x
		\\
		& \leq -\eta x^{2} - \mu_{Q} \Vert z \Vert^{2} - \beta x \varphi(x) + \rho \vert x \vert
		\Vert z \Vert
		\\
		& \leq -\eta x^{2} - \mu_{Q} \Vert z \Vert^{2} - \beta x \varphi(x) + 
		\frac{\rho}{\varepsilon} \vert x \vert^{2} + \rho \varepsilon \Vert z \Vert^{2} \ .
	\end{align*}
	Setting $\varepsilon = \frac{\mu_{Q}}{2 \rho}$ yields
	\begin{displaymath}
		\dot{V}(x,z) \leq -\eta x^{2} - \frac{\mu_{Q}}{2} \Vert z \Vert^{2} 
		- \beta x \varphi(x) + \frac{\rho}{\varepsilon} | x |^2.
	\end{displaymath}
	From \eqref{eq:assum:nonlinear}, there exists $r > 0$ such that,
	for all $\vert x \vert > r$, $\dot{V}(x,z) \leq -\eta x^{2} -
	\frac{\mu_{Q}}{2} \Vert z \Vert^{2}$. Boundedness of solutions
	follows. \hfill $\blacksquare$
\end{pf}
In the next sections we will design particular interconnections 
based on  Theorems \ref{thm:behavior} and \ref{thm:interconnection}.
\subsection{$p$-Passivity and interconnections}
\label{sec:passivity_and_interconnections}
\begin{figure}[htpb]
	\centering
	\includegraphics[width=0.25\textwidth]{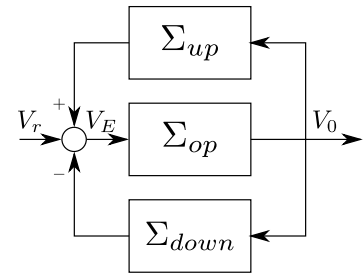}
	\caption{Feedback loops of a circuit with an operational amplifier}
	\label{fig:opampFeedback}
\end{figure}

\begin{figure}[htpb]
	\centering
	\includegraphics{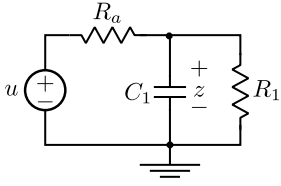}
	\caption{$RC$ network that is both $0$-passive and $1$-passive from different ports}
	\label{fig:simpleRC}
\end{figure}

We consider the interconnection of op-amp \eqref{eq:op-amp}
and linear networks of the form \eqref{eq:linearNetwork}
typically in positive or negative feedback, as show in Figure \ref{fig:opampFeedback}.
Passivity is a theory of negative feedback. In order to apply Theorem \ref{thm:interconnection}
to {\it positive} feedback interconnections, we consider the reverted output $\bar{y}=-y$ and interpret
the  positive feedback interconnection $ 
V_E = + y  + V_r $
 as negative feedback interconnection of the reverted output:
\begin{equation}
\label{eq:pos_feed}
V_E = -\bar{y} + V_r \,, \quad u = V_0 \ .
\end{equation}
We note that  the network in Figure \ref{fig:simpleRC} is
strictly $0$-passive from $u$ to $y=z$ and strictly $1$-passive from
$u$ to $\bar{y}=-z$. 
Indeed, define
${a} = \frac{R_{1}+ R_{a}}{R_{1} R_{a} C_{1}}$, 
${b} = \frac{1}{R_{a}C_{1}}$, and ${c} = 1$.  
Then the transfer functions from  $u$ to $y$ reads
\begin{equation}
	G(s) = \frac{{c} {b}}{s + {a}} 
\end{equation}
whereas the transfer function from $u$ to $\bar{y}$
reads
\begin{equation} 
 \bar{G}(s) = -\frac{{c} {b}}{s + {a}}
\end{equation}
By Theorem \ref{prop:frequency},
$G(s)$ is strictly $0$-passive for any rate $\lambda \in [0, a)$.
On the other hand, $\bar{G}(s)$ is strictly $1$-passive for any  
rate $\lambda \in (a, +\infty)$.
\subsection{$1$-Passive circuits}
\label{sec:1pass}
By Theorem \ref{thm:behavior}, multistable circuits that switch among 
several fixed points may arise from the interconnection of the op-amp 
with strictly $1$-passive networks.
As an illustration, consider the positive feedback of the op-amp
with the $RC$ network in Figure \ref{fig:simpleRC}.
\begin{figure}[htpb]
	\centering
	\includegraphics{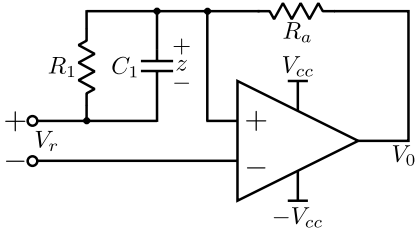}
	\caption{An op-amp in positive feedback with a passive network.}
	\label{fig:bistable}
\end{figure}	
The $RC$ network is strictly $1$-passive from $u$ to $\bar{y}=-z$ 
with rate $\lambda \in (a, +\infty)$,
and takes the role of $\Sigma_{up}$ in Figure 
\ref{fig:opampFeedback}.  
The op-amp is $0$-passive with rate $\lambda \in [0, \frac{1}{R_{0} C_{0}})$.
Thus, for
\begin{equation}
	\frac{1}{R_{0} C_{0}} > a,
	\label{eq:op-amp_fast}
\end{equation}
the two systems share a common interval for their $\lambda$ 
rates\footnote{\eqref{eq:op-amp_fast} requires that the op-amp dynamics is faster than 
the dynamics of the external network, as usual in applications.}. 
By Theorem \ref{thm:interconnection} the 
closed loop system in Figure \ref{fig:bistable} is strictly $1$-passive from
$V_r$ to $V_0$ with
rate $\lambda \in (a, \frac{1}{R_{0} C_{0}})$.
Theorem \ref{thm:boundedness} guarantees boundedness of the 
closed-loop trajectories for any constant input $V_r$. Therefore,
Theorem \ref{thm:behavior} guarantees asymptotic convergence of all
trajectories to some fixed point. In particular, taking $V_r = 0$ for simplicity,
the closed loop is bistable for
\begin{equation}
	\frac{\partial \varphi(x)}{\partial x} \Big|_{x = 0} < -\frac{1}{R_{0} } +
	\frac{\alpha R_{1}}{R_{1} + R_{a}},
	\label{eq:bistable:condition:2}
\end{equation}
which guarantees the existence of three equilibrium points (two stable nodes
and a saddle). 
\subsection{$2$-Passive circuits}
\label{sec:2pass}
\begin{figure}[htpb]
	\centering
	\includegraphics{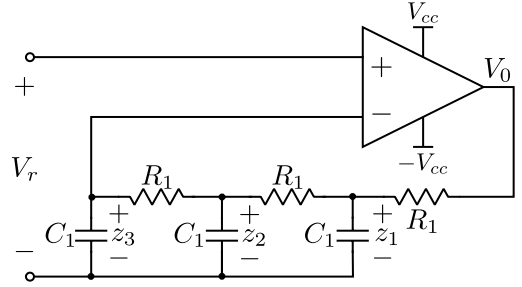}
	\caption{RC oscillator circuit.}
	\label{fig:rcOsc}
\end{figure}

\begin{figure}[htpb]
	\centering
	\includegraphics[width=0.4\textwidth]{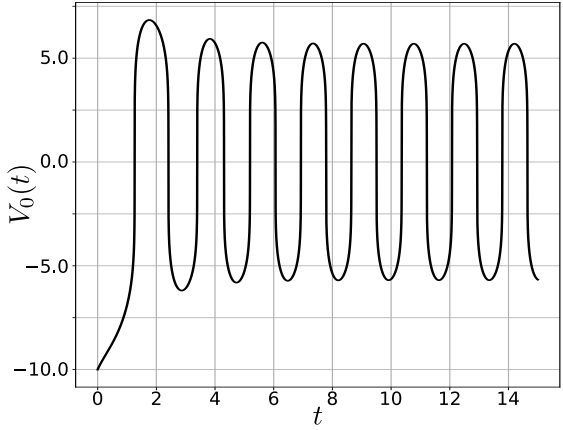}
	\caption{Output of the $RC$ circuit of Figure \ref{fig:rcOsc}, with $R_1 = 3.3 K \Omega$ and
	$C_1 = 200 \mu F$.}
	\label{fig:rcOscOutput}
\end{figure}

By Theorem \ref{thm:behavior}, oscillatory circuits 
may arise from the interconnection of the op-amp 
with strictly $2$-passive networks.
As an illustration, consider the negative feedback of the op-amp
with the $RC$ network in Figure \ref{fig:rcOsc}. 
The $RC$ network admits the state-space representation
$$
	{A} = \frac{1}{R_1C_1}
	\begin{bmatrix}
		-2 & 1 & 0
		\\
		1 & -2 & 1
		\\
		0 & 1 & -1
	\end{bmatrix}, \; {B} = \frac{1}{R_1C_1}
	\begin{bmatrix}
		1 \\ 0 \\ 0
	\end{bmatrix}, \; {C} =
	\begin{bmatrix}
		0 & 0 & 1
	\end{bmatrix}.
$$
The transfer function $G(s)$ from $u$ to $y$ reads
\begin{equation}
	G(s) = \frac{\frac{1}{(R_1 C_1)^{3}}}{s^{3} + \frac{5}{R_1C_1} s^{2} + \frac{6}{(R_1C_1)^{2}}s +
	\frac{1}{(R_1C_1)^{3}}} \ .
	\label{eq:rcNetwork}
\end{equation}
Denoting by $p_{i}$ the $i$-th pole of $G$ and by $\beta_{i} = \vert \Re\{p_{i}\} \vert$ the magnitude of the real part of the poles of $G$
(without loss of generality, we assume $0 \leq \beta_{1} \leq \beta_{2} < \beta_{3}$),
the $RC$ network is strictly $2$-passive from $u$ to $y$ with rate 
$\lambda \in \left( \max \left\{ \beta_{2}, \frac{\beta_{1} + \beta_{2} + \beta_{3}}{3} \right\} , \beta_3\right)$.
The network is thus constrained to a negative feedback interconnection, taking the role of $\Sigma_{down}$ in Figure \ref{fig:opampFeedback}.  
For
\begin{equation}
\label{eq:cond:2passive}
	\frac{1}{R_{0} C_{0}} > \max \left\{ \beta_{2}, \frac{\beta_{1} + \beta_{2} + \beta_{3}}{3} \right\} 
\end{equation}
the op-amp and the RC network share a common interval for their $\lambda$ 
rates. By Theorem \ref{thm:interconnection} the 
closed loop system in Figure \ref{fig:rcOsc} is thus strictly $2$-passive from
$V_r$ to $V_0$ with rate $\lambda \in \left (\max \left\{ \beta_{2}, \frac{\beta_{1} + \beta_{2} + \beta_{3}}{3} \right\} , \min \left\{\beta_3, \frac{1}{R_{0} C_{0}}\right\}\right)$.
Theorem \ref{thm:boundedness} guarantees boundedness of the 
closed loop trajectories for any constant input $V_r$, thus 
Theorem \ref{thm:behavior} guarantees that the trajectories of the four
dimensional closed-loop circuit all converge to a simple attractor. 
For $R_1 = 3.3 K \Omega$ and $C_1 = 200 \mu F$, $G(s)$ has poles
$p_{1} = -0.3$, $p_{2} = -2.35$ and $p_{3} = -4.92$. Therefore, strict $2$-passivity
holds for $\lambda \in (2.52, 4.92)$. For op-amp parameters
$\alpha = 0.1$, $R_{0} = 1 M \Omega$, $C_{0} = 15.9 nF$, 
$\varphi(x) = (\frac{x}{12})^{5}$ condition \eqref{eq:cond:2passive} holds.
These specific parameters also ensure that the unique fixed point at the origin is unstable. 
Thus, by Theorem \ref{thm:behavior},
every trajectory converges asymptotically to a limit cycle. The steady-state
is an oscillation, as shown in Figure \ref{fig:rcOscOutput}.
\section{Mixed feedback and modulation}
\label{sec:mixedfeedback_and_modulation}
\begin{figure}[htpb]
	\centering
	\includegraphics[width=0.64\columnwidth]{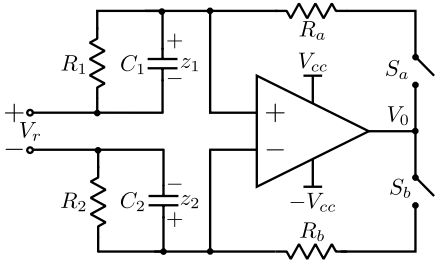}
	\caption{Mixed feedback with switches.}
	\label{fig:mixed:switch}
\end{figure}

\begin{figure}[htpb]
	\centering
	\includegraphics[width=0.58\columnwidth]{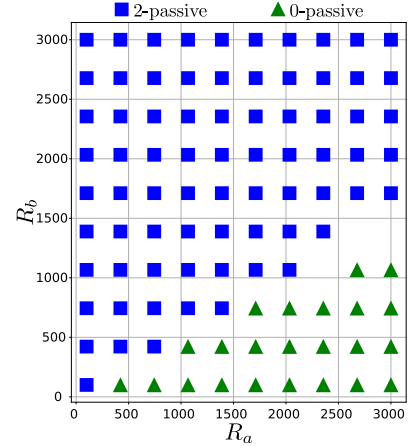} \vspace{-2mm}
	\caption{$p$-passivity of the circuit of Figure \ref{fig:mixed:switch}  with both switches closed,
	as a function of the resistors $R_{a}$ and $R_{b}$. Gaps correspond to lack of $p$-passivity.}
	\label{fig:modulation2}
\end{figure}

\begin{figure}[htpb]
	\centering
	\includegraphics[width=0.4\textwidth]{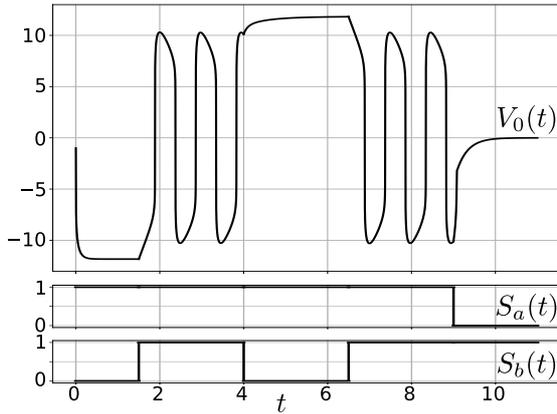} \vspace{-2mm}
	\caption{Transitions among different behaviors of circuit of Figure \ref{fig:mixed:switch} 
	driven by the switches $S_{a}$ and $S_{b}$; 0 - open switch, 1 - closed switch. 
	Recall that  
	for $(S_{a},S_{b}) = (0,1)$ the circuit is $0$-passive,
	for $(S_{a},S_{b}) = (1,0)$ the circuit is $1$-passive, and 
	for $(S_{a},S_{b}) = (1,1)$ the circuit is $2$-passive.}
	\label{fig:modulation1:output}
\end{figure}

The mixed feedback amplifier is a classical device of nonlinear circuit theory \citep{chua1987}.
It combines positive and negative feedback around an operational amplifier to create nonlinear behaviors.
We illustrate this flexibility with  the simple system of Figure \ref{fig:mixed:switch}.
The two linear networks correspond to two copies of the $RC$ network in Figure \ref{fig:simpleRC}.
The behavior of the closed loop is dictated by the interconnection pattern of the switches $S_{a}$ and 
$S_{b}$. If $S_{a}$ is closed and $S_{b}$ is open then the network reduces to the one in
Figure \ref{fig:bistable}; the closed loop is $1$-passive. If $S_{a}$ is open and 
$S_{b}$ is closed, then the closed loop is $0$-passive. When both switches are closed, 
the feedback circuit is not necessarily $1$-passive 
because the rates of the two networks may not be compatible. In fact, 
a suitable selection of the network parameters lead to richer behaviors.
Take $S_{a}$ and $S_{b}$ both closed and define 
 $a_{1} = \frac{1}{R_{a} C_{1}}$, $a_{2} = \frac{1}{R_{b} C_{2}}$, and $b_{i} =a_{i} + 
\frac{1}{R_{i}C_{i}}$, $i = 1,2$. With these data, the upper network is strictly $1$-passive from $V_0$ to
$-z_1$ and the lower network is strictly $0$-passive from $V_0$ to $z_2$, respectively with rates
$\lambda_{up} \in (b_1,+\infty)$ and $\lambda_{down} \in [0,b_2)$. 
A common rate $\lambda$ can be found for $b_{1} < \min \{ b_{2}, \frac{1}{R_{0} C_{0}} \}$,
since the op-amp is $0$-passive with rate $\lambda_{op} \in [0, \frac{1}{R_{0} C_{0}})$.
In this case, the feedback circuit is $1$-passive, by Theorem  \ref{thm:interconnection}. 
In contrast, Theorem  \ref{thm:interconnection} cannot be used for $b_1 > b_2$ because
of the absence of a common rate. However, the aggregate transfer function reads
\begin{equation}
	G(s) = \frac{(a_{2} - a_{1})s + a_{2} b_{1} - a_{1} b_{2}}{(s + b_{1})(s + b_{2})},
	\label{eq:G_mixed}
\end{equation}
which has positive real part if and only if there exist $\lambda \in [0, b_{2}) \cup (b_1, +\infty)$ such that 
\begin{align}
 	(a_{2} - a_{1}) \lambda & < a_{2} b_{2} - a_{1} b_{1}
	\label{eq:condition:mixed:2}
	\\
	(a_{2} - a_{1}) \lambda & < a_{2} b_{1} - a_{1} b_{2}.
	\label{eq:condition:mixed:3}
\end{align}
Indeed,
\begin{enumerate}
\item 
	$G(s)$ is strictly $0$-passive\footnote{Indeed, since $b_{1} > b_{2}$, it follows that $a_{1} b_{1} > a_{1} b_{2}$
	and $a_{2} b_{1} > a_{2} b_{2}$. Hence, $a_{2} b_{2} -a_{1} b_{1} < a_{2}b_{1} - a_{1}b_{2}$
	and conditions \eqref{eq:condition:mixed:2} and \eqref{eq:condition:mixed:3} reduce to 
	$\lambda < \frac{a_{2} b_{2} - a_{1} b_{1}}{a_{2} -a_{1}}$. Moreover, since $b_{1} > b_{2}$ it follows that
	$\lambda < b_{2}$. From this last observation, together with 
	$\lambda \in [0, b_{2}) \cup (b_1, +\infty)$ and Proposition \ref{prop:frequency} it follows that $G(s)$ is $0$-passive.}
	with rate $\lambda \!\in\! \left[ 0, \frac{a_{2} b_{2} - a_{1} b_{1}}{a_{2} -a_{1}} \right)$ for $a_{2} > a_{1}$  and
	$\frac{a_{2} b_{2} - a_{1} b_{1}}{a_{2} -a_{1}} > 0$;
\item 
	$G(s)$ is strictly $2$-passive with rate $\lambda \!\in\! \left( \!\frac{a_{2} b_{2} - a_{1} b_{1}}{a_{2} - a_{1}}, +\infty \!\right)$ for
	$a_{1} > a_2$. 
\end{enumerate}
The conditions above reveal that mixed feedback allows for both $0$-passivity and $2$-passivity.
The network behavior can be modulated from monostable to oscillatory
via parameter variations. Figure \ref{fig:modulation2} shows the degree of 
$p$-passivity of the closed loop for different values $0 \Omega < R_{a}, R_{b} < 3K\Omega$.
Indeed, transitions from monostable to oscillatory regimes are obtained by the
variation of one of the two resistances. 
As a final illustration, we consider parameters 
$R_{1} = R_{2} = 3.3 K \Omega$, $R_{a} = R_{b} = 1 K \Omega$, $C_{1} = 100 \mu F$, 
$C_{2} = 200 \mu F$, $R_{0} = 1 M \Omega$, $C_{0} = 15.9 n F$ and $\alpha = 1$.
For these parameters $G(s)$ in \eqref{eq:G_mixed} is strictly $2$-passive. 
When both switches are closed the origin is the only equilibrium point and is unstable. Hence, 
by Theorems \ref{thm:behavior} and \ref{thm:boundedness} we conclude the existence of a limit cycle. 
Figure \ref{fig:modulation1:output} shows transitions among different
behaviors, driven by the switches.
\section{Conclusions}
We have analyzed basic examples of circuits that switch and oscillate as interconnections
of linear circuits and operational amplifiers. The approach builds upon 
dominance theory and $p$-passivity. The saturated op-amp model  
guarantees boundedness of trajectories in closed loop and allows
for positive and negative feedback interconnections with linear
$p$-passive networks. Specific interconnections lead to $1$- and $2$-passive
networks, leading to a tractable analysis of bistability and  oscillations in possibly high-dimensional
models.
The stability analysis in this paper is based on solving linear matrix inequalities
very much as in the stability analysis of linear systems. Such a computational
framework suggests many possible extensions to analyze the performance
and robustness of switching and oscillatory circuits in the same way as
for linear systems.  
 
\bibliography{Biblio.bib,science}             

\begin{thebibliography}{18}
\providecommand{\natexlab}[1]{#1}
\providecommand{\url}[1]{\texttt{#1}}
\providecommand{\urlprefix}{URL }
\expandafter\ifx\csname urlstyle\endcsname\relax
  \providecommand{\doi}[1]{doi:\discretionary{}{}{}#1}\else
  \providecommand{\doi}{doi:\discretionary{}{}{}\begingroup
  \urlstyle{rm}\Url}\fi

\bibitem[{Aghannan and Rouchon(2003)}]{aghannan2003}
Aghannan, N. and Rouchon, P. (2003).
\newblock An intrinsinc observer for a class of {Lagrangian} systems.
\newblock \emph{IEEE Transactions on Automatic Control}, 48(6).

\bibitem[{Chua et~al.(1987)Chua, Desoer, and Kuh}]{chua1987}
Chua, L.O., Desoer, C.A., and Kuh, E.S. (1987).
\newblock \emph{Linear and nonlinear circuits}.
\newblock Mc-Graw Hill.

\bibitem[{Clayton and Winder(2003)}]{clayton2003operational}
Clayton, G. and Winder, S. (2003).
\newblock \emph{Operational Amplifiers}.
\newblock EDN Series for Design Engineers. Elsevier.

\bibitem[{Forni and Sepulchre(2013)}]{forni2013b}
Forni, F. and Sepulchre, R. (2013).
\newblock On differentially dissipative dynamical systems.
\newblock In \emph{9th IFAC Symposium on Nonlinear Control Systems}, 15--20.
  Toulouse, France.

\bibitem[{Forni and Sepulchre(2014)}]{forni2014b}
Forni, F. and Sepulchre, R. (2014).
\newblock A differential {Lyapunov} framework for contraction analysis.
\newblock \emph{IEEE Transactions on Automatic Control}, 59(3), 614--628.

\bibitem[{Forni and Sepulchre(2017{\natexlab{a}})}]{Forni2017b}
Forni, F. and Sepulchre, R. (2017{\natexlab{a}}).
\newblock Differential dissipativity theory for dominance analysis.
\newblock Submitted to IEEE Transactions on Automatic Control,
  http://arxiv.org/abs/1710.01721.

\bibitem[{Forni et~al.(2013)Forni, Sepulchre, and van~der Schaft}]{forni2013}
Forni, F., Sepulchre, R., and van~der Schaft, A.J. (2013).
\newblock On differential passivity of physical systems.
\newblock In \emph{52rd IEEE Conference on Decision and Control}, 6580--6585.
  Florence, Italy.

\bibitem[{Forni and Sepulchre(2017{\natexlab{b}})}]{forni2017}
Forni, F. and Sepulchre, R. (2017{\natexlab{b}}).
\newblock A dissipativity theorem for $p$-dominant systems.
\newblock In \emph{56th IEEE Conference on Decision and Control}. Melbourne,
  Australia.

\bibitem[{Jouffroy and Fossen(2010)}]{jouffroy2010}
Jouffroy, J. and Fossen, T.I. (2010).
\newblock A tutorial on incremental stability analysis using contraction
  theory.
\newblock \emph{Modeling, Identification and Control}, 31(3), 93--106.

\bibitem[{Karki(2000)}]{karki2000}
Karki, J. (2000).
\newblock Effect of parasitic capacitance in op amp circuits.
\newblock Technical Report SLOA013A, Texas Instruments.

\bibitem[{Lohmiller and Slotine(1998)}]{lohmiller1998}
Lohmiller, W. and Slotine, J.J.E. (1998).
\newblock On contraction analysis for nonlinear systems.
\newblock \emph{Automatica}, 34(6), 683--696.

\bibitem[{Miranda-Villatoro et~al.(2017)Miranda-Villatoro, Forni, and
  Sepulchre}]{Miranda2017b}
Miranda-Villatoro, F.A., Forni, F., and Sepulchre, R. (2017).
\newblock Analysis of {L}ur'e dominant systems in the frequency domain.
\newblock Submitted to Automatica, http://arxiv.org/abs/1710.01645.

\bibitem[{Miranda-Villatoro et~al.(2018)Miranda-Villatoro, Forni, and
  Sepulchre}]{miranda2018}
Miranda-Villatoro, F.A., Forni, F., and Sepulchre, R. (2018).
\newblock Dominance analysis of linear complementarity systems.
\newblock Submitted to MTNS 2018, http://arxiv.org/abs/1802.00284.

\bibitem[{Noseek(2009)}]{noseek2009}
Noseek, J.A. (2009).
\newblock Circuit elements, modeling and equation formulation.
\newblock In \emph{The Circuits and Filters Handbook: Fundamentals of Circuits
  and Filters}, 13--1, 13--11.

\bibitem[{Pavlov and L(2008)}]{pavlov2008}
Pavlov, A. and L, M. (2008).
\newblock Incremental passivity and output regulation.
\newblock \emph{Systems \& Control Letters}, 57, 400--409.

\bibitem[{Pavlov et~al.(2005)Pavlov, Van De~Wouw, and Nijmeijer}]{pavlov2005}
Pavlov, A., Van De~Wouw, N., and Nijmeijer, H. (2005).
\newblock Convergent systems: analysis and synthesis.
\newblock In \emph{Control and observer design for nonlinear finite and
  infinite dimensional systems}, 131--146. Springer.

\bibitem[{Stan and Sepulchre(2007)}]{stan2007}
Stan, G.B. and Sepulchre, R. (2007).
\newblock Analysis of interconnected oscillators by dissipativity theory.
\newblock \emph{IEEE Transactions on Automatic Control}, 52(2), 256--270.

\bibitem[{van Der~Schaft(2013)}]{schaft2013}
van Der~Schaft, A.J. (2013).
\newblock On differential passivity.
\newblock In \emph{9th IFAC Symposium on Nonlinear Control Systems}, 21--25.
  Toulouse, France.

\end{thebibliography}
\end{document}